\documentstyle[prb,multicol,aps,epsfig]{revtex}

\begin{document}

\title{Orbital Ordering in Paramagnetic LaMnO$_3$ and KCuF$_3$.}

\author{
J.~E. Medvedeva\(^{1,2}\), M.~A. Korotin\(^{1}\),
V.~I. Anisimov\(^{1}\), A.~J. Freeman\(^{2}\)}

\address{ 	
\(^1\) Institute of Metal Physics, Yekaterinburg, Russia. \\
\(^2\) Dept. of Physics and Astronomy, Northwestern, 
Evanston, Illinois 60208-3112. 
}

\maketitle

\begin{abstract}
{\it Ab-initio} studies of the stability of orbital ordering,
its coupling to magnetic structure and its possible origins
(electron-phonon and/or electron-electron interactions)
are reported for two perovskite systems, LaMnO$_3$ and KCuF$_3$.
We present a new average spin state calculational scheme
that allowed us to treat a paramagnetic state
and to succesfully describe the experimental
magnetic/orbital phase diagram of both LaMnO$_3$ and KCuF$_3$
in crystal structures when the Jahn-Teller distortions are neglected.
Hence, we conclude that the orbital ordering in both compounds 
is purely electronic in origin. 
\end{abstract}

\begin{multicols}{2}

It is known from earlier studies \cite{Goodenough55,KugelKhomskii} 
that magnetic ordering is coupled to the orbital ordering (OO) -- 
magnetic interactions depend on the type of occupied orbitals of
transition-metal ions. (A very detailed overview on magnetic and 
orbital ordering (OO) in cuprates and manganites was given recently
by Ole{\'s} {\it et al} \cite{Oles}.)
For example, the OO in LaMnO$_3$ reduces the ferromagnetic (FM)
contribution from the $e_g$ orbitals to the interlayer exchange coupling
and is responsible for the stability of the magnetic ground state structure
(which is A-type antiferromagnetic (AFM) below T$_N$=141 K \cite{Wollan55}) 
--- as described both experimentally and theoretically
on this manganite \cite{Goodenough55,Wollan55,Solovyev99}.
In KCuF$_3$, the OO was obtained in a spin-orbital model 
\cite{KugelKhomskii} as a result of an exchange interaction which correctly
describes both orbital and spin alignments simultaneously;
thus the spin and orbital degrees of freedom are mutually coupled.

A more complicated picture of the fundamental connection between
OO phenomena and forms of the spin alignment
follows from experimental measurements that characterize an OO
parameter and magnetic interactions in transition-metal oxides.
Using the dipole resonant x-ray scattering technique, Murakami
{\it et al.} \cite{Murakami98} found a sharp disappearence of
OO at much higher temperatures ($\sim$780 K) than T$_N$
in LaMnO$_3$. They suggested a coupling of the spin
and orbital degrees of freedom due to the small decrease of the order
parameter of the orbital structure above T$_N$, even despite
the presence of OO in the paramagnetic (PM) phase of LaMnO$_3$.
The most striking result was obtained for bilayered LaSr$_2$Mn$_2$O$_7$ 
\cite{Chatterji00}, namely,
a competition between the A-type AFM spin ordering
with T$_N \approx $170 K and the CE-type \cite{Wollan55}
charge/orbital ordering (COO), which exists between $T_N$ and
COO transition temperature, T$_{COO}$=210 K.
With decreasing temperature, the development of
the COO phase is disrupted by the onset of
the A-type AFM ordering at T$_N$.

In this paper, we investigate the stability of observed OO's
without appealing to magnetic interactions by modeling
the PM phase of two manganites, LaMnO$_3$ and KCuF$_3$.
To this end, we examined all magnetic configurations (MC's) that are
mathematically possible in a unit cell of the perovskites
which consists of 4 Mn (or Cu) atoms. There are only eight such
spin sets -- as schematically presented in Fig. 2(a).
They include the possibility of having FM (labelled ''1''),
and C-type (''3''), A-type (''6'') and G-type (''7'') AFM 
spin orderings
-- as well as four MC's that have a mathematical meaning,
but cannot be realized as an individual magnetically ordered
solution.
Then, to model a PM state, during the iterations towards 
self-consistency, we average the orbital occupation matrices (that
contain information on the orbital polarization and OO in the unit cell)
and potential parameters of all atoms over eight 
spin alignment configurations after each iteration.
As a result, we have an averaged spin state and can analyze
the stability/presence of orbital polarization in such a modeled
PM phase. Self-consistency of the averaged spin state calculation (ASSC)
is determined by convergence of the averaged total energy, charge
densities and orbital occupation matrices.

Typically, the origin of the orbital polarization is believed to be
the electron-phonon interaction (the Jahn-Teller (JT) distortion)
and the electron-electron interaction.
A cooperative JT effect which sets in well above the magnetic
transition temperature, stabilizes the particular order of the partly filled
$e_g$ orbitals that are close to orbital degeneracy \cite{KugelKhomskii}.
Monte Carlo investigations \cite{Hotta99} showed a
stabilization of the experimental magnetic and OO
in LaMnO$_3$ only after including lattice distortions.
However, a spin-orbital model for insulating undoped LaMnO$_3$
derived in \cite{Feiner99}
qualitatively explains the observed A-type AFM ordering
as stabilized by a purely electronic mechanism.

We investigated two typical pseudocubic perovskites:
{\it (i)} LaMnO$_3$, in which the strong JT effect
breaks the degeneracy of the electronic configuration of Mn$^{3+}$
($t_{2g}^3e_g^1$) and directly affects the $e_g$ orbital population.
A recent neutron-powder diffraction study,
together with thermal analysis \cite{Rodriguez98}, showed that LaMnO$_3$
undergoes a structural phase transition at T$_{JT}$=750 K above which
the OO disappears. To establish the origin of OO
in LaMnO$_3$, we performed ASSC's for the high temperature 
crystal structure where JT distortions are absent. 
Since diffraction data \cite{Rodriguez98,Huang97}
can be interpreted using two crystal symmetries --
double-cubic orthorhombic $Pbnm$ structure  and rhombohedral
$R\overline{3}c$ crystal structure -- we did calculations for both.
{\it (ii)} Another example of a perovskite compound with partly occupied
$e_g$ orbitals resulting in strong JT distortions
and strong one-dimensional AFM interactions is KCuF$_3$.
There are two tetragonal crystal types
of KCuF$_3$ \cite{Hutchings69+Okazaki69}, which are stable
in a wide temperature range, and their A-type AFM ordering
temperatures are equal to 20 K and 38 K.
It was shown \cite{Liecht95} for this compound
that the ground state with
the lattice distortion and the orbital polarization (and ordering)
can be stabilized only by taking into account the Coulomb correlation
between d-shell electrons (LDA+U approach \cite{ldau}).
We use Coulomb and exchange parameters, U=8 eV and J=0.88 eV 
for both compounds as those relevant to the 3d transition metal
oxides \cite{ldau,satpathy96+solovyev96}.

{\it LaMnO$_3$.}
As the first step, we performed LDA+U calculations for the experimental
A-type AFM phase of LaMnO$_3$ with low-T $Pbnm$ 
structure (where strong JT distortions are present).
The self-consistent non-diagonal occupation matrix for the spin density
of the $e_g$ subshell ($3z^2-r^2$ and $x^2-y^2$ orbitals) of one
particular Mn atom is found to be
$$ n^{\uparrow}_{mm'} - n^{\downarrow}_{mm'} =
\left( \begin{array}{cc}
0.43 & 0.29 \\
0.29 & 0.59 \\
\end{array} \right). $$
Its diagonalization gives two new $e_g$ orbitals:
$\phi _1=3x^2-r^2$ ($3y^2-r^2$ for the second type of Mn atom)
with an occupancy of 0.81
and $\phi _2=z^2-y^2$ ($z^2-x^2$) with an occupancy of 0.20.
Since a Mn$^{3+}$ ion has formally one electron in the partially filled
$e_g^{\uparrow}$ subshell and the $t_{2g}^{\uparrow}$ subshell is
totally filled, we presented the resulting OO
by plotting the angular distribution of the $e_g$ spin density
in Fig. 1(a): 
$ \rho(\theta ,\phi)=\sum_{mm'}(n_{mm'}^\uparrow - n_{mm'}^\downarrow)
Y_m(\theta,\phi)Y_{m'}(\theta ,\phi) $,
where the $Y_m(\theta, \phi)$ are corresponding spherical harmonics.

The resulting picture of the OO in A-type AFM
LaMnO$_3$ is in very good agreement with that given by Goodenough
\cite{Goodenough55} and recently detected
using resonant x-ray scattering \cite{Murakami98}.
Based on this result, we can now move to the case of $T>T_N$
to investigate the region of the magnetic/orbital phase diagram
where the OO exists in the PM phase \cite{Murakami98}.
To this end, we performed the model ASSC's described above
for the high-T (at T=798 K) structure of LaMnO$_3$
with experimental lattice parameters \cite{Rodriguez98}.
The PM iterations were started from a
uniform distribution of $e_g$ electrons over the $3z^2-r^2$ and $x^2-y^2$
orbitals of all Mn atoms.
In Fig. 1(b), we plot the "total energy" of the PM state
of the high-T structure of LaMnO$_3$ (crosses)
obtained by averaging the total energies of the eight
possible MC's.
This total energy curve becomes saturated after the $\sim$20th iteration;
hence, we may call this ASSC result self-consistent.

To illustrate the stability of OO in the PM
phase and to compare the orbital polarization with that calculated
for the experimental A-type AFM phase, we present the OO
as a difference between the occupancies of the (''diagonalized'')
$\phi _1$ and $\phi _2$ orbitals.
In Fig. 1, this difference is shown plotted against the number
of PM iterations (crosses).
We found that starting with equal $3z^2-r^2$ and $x^2-y^2$ orbital
occupancies (and with a diagonal occupation matrix), 
the value of the non-diagonal
elements of the occupation matrix grows during the ASSC
leading to a saturated orbital polarization of the right type (the same
obtained for A-type AFM LaMnO$_3$, Fig. 1(a))
around the tenth iteration (Fig. 1).
Note that the OO in the PM state is the only stable solution.
Forcing different orbital polarizations (ferro-orbitally ordered, etc.)
as starting ones for the PM calculation resulted in
the type of OO obtained above by changing the corresponding
(non-diagonal) orbital occupancy during the iterations towards 
self-consistency.
The horizontal line in Fig. 1 stands for the difference between $\phi _1$
and $\phi _2$ orbital occupancies obtained from the usual spin-polarized
self-consistent LDA+U calculation
for the A-type AFM phase of LaMnO$_3$ in the low-T $Pbnm$ structure.
Comparing the orbital polarizations of the PM and AFM
phases (Fig. 1), we conclude that the AFM-PM transition in LaMnO$_3$
results in a slight decrease of the difference between the $\phi _1$
and $\phi _2$ orbital occupancies but does not supress the OO.
This finding
agrees with the observed behavior of the OO parameter
when T increases ($T>T_N$) \cite{Murakami98}.

We need to point out here one essential feature of the ASSC.
Since we consider a spin state averaged over eight MC's
and therefore an average ocupation matrix of such a PM state,
it is interesting to analyze the deviation of the orbital polarization
in each of these MC's from the average value.
As seen from Fig. 3(b), where we plot the $\phi _1$ and $\phi _2$ 
orbital occupancy difference for each of the eight MC's
after the first (broken line) and fiftieth (solid line)
PM iteration, the deviations from the average value
become smaller with iterations -- they decrease from 36\% to 12\%,
correspondingly. (Further increases of the iteration number
do not supress the deviations.)
The variations at the beginning of the ASSC and for the result obtained,
c.f., Fig. 2(b), have a close analogy in oscillation behavior, i.e.,
the location of the $e_g$ orbital occupancy difference
stays higher (lower) with respect to the average value
for the same particular MC during the iterations.
Hence, one can separate the eight MC's into three
groups: the lowest (MC's 1 and 2), the highest (4 and 7) and those close
to the average value (3,5,6 and 8).

We analyzed possible reasons for this separation by calculating
the number of FM or AFM neighbours aligned
to the first Mn (whose spin does not change in all eight MC's, Fig. 2(a)).
First, we found that the behavior of the $e_g$ orbital
occupancy difference curves (Fig. 2(b))
is determined by the nearest neighbours' spin alignment.
For both the 1st and 2nd MC's,
all six nearest neighbours of the first Mn
have the same spin direction, so FM
couplings with all nearest neighbours give the smallest
$\phi _1-\phi _2$ value. By contrast, the largest contributions
were obtained from MC's 4 and 7, which have six
AFM couplings between the first Mn and
its nearest neighbors. The third group of MC's
possesses two FM and four AFM couplings
(MC 3 and 8), giving almost an exact
average value $\phi _1-\phi _2$, and four FM
and two AFM couplings (MC 5 and 6), both lying below the average value.
(Note here, that it would be sufficient to consider only four
MC's: the 1st (which corresponds to the FM phase), the
3rd (C-type AFM), the 6th (A-type AFM) and the 7th (G-type AFM).
Averaging over these four MC's underestimates the value averaged
over all eight MC's by only 0.7 and 0.1 \% at the
first and 50th PM iteration, respectively.)
Thus, we can conclude that the AFM spin alignment
seems to be more preferable for OO than the FM one;
however, the decrease of the $\phi _1-\phi _2$ deviation against
the average values (from 47 to 17 \% for the first and 50th iteration,
respectively) in the FM phase also supports the importance
of the FM coupling in ASSC.

Let us now discuss some peculiarities in the crystal structure
of LaMnO$_3$ related to the JT distortions.
As found by recent neutron-powder diffraction studies
\cite{Rodriguez98}, the JT transition in this perovskite
compound occurs at T=750 K.
If we define the degree of tetragonal distortion as
$ \delta_{JT} = (d_l - d_s)/(d_l+d_s)/2 $,
where $d_l$ and $d_s$ denote the long and short Mn-O bond distances,
then according to the structural data \cite{Rodriguez98},
$\delta_{JT}$ changes with heating from $\delta_{JT}$=0.133 for 
the room-T $Pbmn$ structure to $\delta_{JT}$=0.023 for
the crystal structure at 798 K (which is also orthorhombic $Pbmn$,
but described as "double-cubic" perovskite \cite{Vegas86}).
Based on the results obtained above, we may conclude that
strong lattice distortions do not influence the OO.
However, $\delta_{JT}$ is not zero in the high-T structure
and we have slightly different Mn-O bond distances
which break the cubic symmetry, and
so the JT effect may still play some role in OO.
To check this possibility, we performed ASSC
for the rhombohedral $R\overline{3}c$ structure with lattice parameters
taken from \cite{Mitchell96}. All Mn-O distances are equal
in this structure, so $\delta_{JT}$=0.000. For this case, the calculated
difference between occupancies of the $\phi _1$ and $\phi _2$ orbitals
against the number of PM
iterations is shown in Fig. 1 (circles). This difference is smaller
than those obtained for the two orthorhombic structures; nevertheless,
we conclude that the OO does not disappear in regular
MnO$_6$ octahedra, and hence JT-distortions are not the origin of 
the OO in LaMnO$_3$.

{\it KCuF$_3$.}
As shown by Kugel and Khomskii \cite{KugelKhomskii},
KCuF$_3$ is an example of a system in which
the exchange interaction alone results in the correct OO.
{\it Ab-initio} LDA+U investigations \cite{Liecht95}
confirmed the electronic origin of the ordering:
the coupling to the lattice is not a driving force for the orbital
(and magnetic) ordering, but the lattice follows the orbital state.
Hence, we performed ASSC's for a model structure
of KCuF$_3$ in which cooperative JT lattice distortions were neglected.
We used the tetragonal ($P4/mmm$) crystal structure with $a$=5.855 \AA,
$c$=7.846 \AA \, and the coordinates: 
K (0, 0.5, 0.45), Cu (0, 0, 0), F1 (0, 0, 0.45) and
F2 (0.25, 0.25, 0). In this structure, the  CuF$_6$ octahedra
are slightly compressed along the $c$ axis: the distance from the Cu
atom to the apical F1 atom
D(Cu-F1)=1.96 \AA, while in the $ab$-plane D(Cu-F2)=2.07 \AA  \,
(without the quadrupolar deformation in the $ab$ plane
that is present in the experimental structure).

Since this perovskite compound possesses
two energetically equivalent types of OO's characterized
by alternation of the $x^2-z^2$ and $y^2-z^2$ orbitals
\cite{KugelKhomskii,Hirakawa71}, we performed four ASSC's
which started with different $e_g$-electron distributions
over Cu $e_g$ orbitals: the orbital polarizations chosen
corresponded to F-type ferro-orbital and A-, C- or G-type antiferro-orbital
alternations of $x^2-z^2$/$y^2-z^2$ orbitals (shown in the upper
row of OO's in Fig. 3); the C- and G- orbital
configurations were observed experimentally \cite{Hirakawa71}.
Since a Cu$^{2+}$ ion ($d^9$ configuration) has one hole in the
$e_g$ state,
we again represent the OO's by a 2$\times$2 $e_g$-orbital
occupation matrix, $n_{mm'}^\uparrow - n_{mm'}^\downarrow$.
As a start, we used
$$ n^{\uparrow}_{mm'} - n^{\downarrow}_{mm'} =
\left( \begin{array}{cc}
0.53 & \pm 0.28 \\
\pm 0.28 & 0.15 \\
\end{array} \right), $$
whose diagonalization gives $x^2-z^2$ ($y^2-z^2$) orbitals.

In contrast to LaMnO$_3$, we found some stable solutions
with different OO's in KCuF$_3$.
In Fig. 3, we present the total energy of the PM phase
for each of the orbital ordered states obtained against
the number of PM iterations. The
corresponding angular distributions of the $e_g$ spin density
for every solution (at the 200th iteration) are shown (bottom row
of OO's in Fig. 3); the antiferro-orbitally ordered state (A) has
the highest total energy. Although it seemed to be stabilized with
the lowest total energy and with almost fully occupied $3z^2-r^2$
orbitals at all Cu sites at around the 70th PM iteration, 
further PM iterations resulted in a sharp increase of the total energy 
of this A-type solution, and at the 200th iteration the OO is an alternation 
of $3z^2-r^2$ and $3x^2-r^2$/$3y^2-r^2$ orbitals.
For the calculation, which was started from F-type ferro-orbital ordering,
the same $x^2-y^2$ orbitals at all Cu sites (upper row of Fig. 3)
change to almost fully occupied $3z^2-r^2$ orbitals
during the PM iterations (bottom row in Fig. 3).

More preferable in energy are two (almost degenerate) solutions
which correspond to G- and C-type OO. Their $\phi _1$ and $\phi _2$ 
orbitals, $x^2-z^2$ and $y^2-z^2$, stagger in the $ab$ plane, 
but have ferro- or antiferro-orbital alignments along the $c$ axis for
C or G-type ordering, respectively. This result agrees
with previous theoretical and experimental studies on OO
in KCuF$_3$ \cite{KugelKhomskii,Liecht95,Hirakawa71}.
Thus, from the behavior of the total energy curves of these four
solutions, we conclude that PM KCuF$_3$ with uniform
CuF$_6$ octahedra possesses two OO's of the right type.

In summary, the new ASSC scheme presented to treat a PM state 
succesfully described the experimental magnetic/orbital phase 
diagram of both LaMnO$_3$ and KCuF$_3$ without inclusion of
JT distortions. The OO in the PM phase is found to be
the same type as in the AFM phase. A small decrease
of the order parameter of the OO upon the AFM-PM transition
is in agreement with the observed one.
Finally, we conclude that OO in both LaMnO$_3$ and KCuF$_3$
is of purely electronic origin. 

Work supported by the Russian Foundation for Basic Research
(grant RFFI-01-02-17063) and the U.S. Department of Energy 
(grant No. DE-F602-88ER45372).


\begin{figure}
\caption{
Dependence of the $\phi _1 - \phi _2$ difference on the number 
of PM iterations for the high-T PM phase,
$\delta_{JT}$=0.023 (crosses) and $R\overline{3}c$ PM phase,
$\delta_{JT}$=0.000 (circles) of LaMnO$_3$.
The horizontal line stands for calculated OO
of $Pbmn$ A-type AFM phase, $\delta_{JT}$=0.133.
In the insert, {\it (a)} denotes the angular distribution 
of the $e_g$-electron spin density in the $Pbmn$ A-type AFM phase 
of LaMnO$_3$ from the LDA+U calculation and
{\it (b)} the total energy averaged over all MC's
as a function of the number of PM iterations
for the two PM phases.
}
\end{figure}

\begin{figure}
\caption{
{\it (a)} Schematic illustration of all possible MC's
in the unit cell which contains four formula units;
only the $ac$ plane is shown. 
{\it (b)} Deviations of the $\phi _1 - \phi _2$  difference 
in each of the MC's considered 
from the averaged value as a result of ASSC's of 
the high-T PM phase of LaMnO$_3$.
Only the first (broken line) and fiftieth (solid line) 
iterations are shown.
}
\end{figure}

\begin{figure}
\caption{
Total energies of four solutions with different OO's
for KCuF$_3$ as a function of the number of PM iterations.
The corresponding OO (F-type ferro-orbital and
A-, C- or G-type antiferro-orbital orderings) are shown
for the zeroth (top row) and the 200th iteration (bottom row).
}
\end{figure}

\end{multicols}
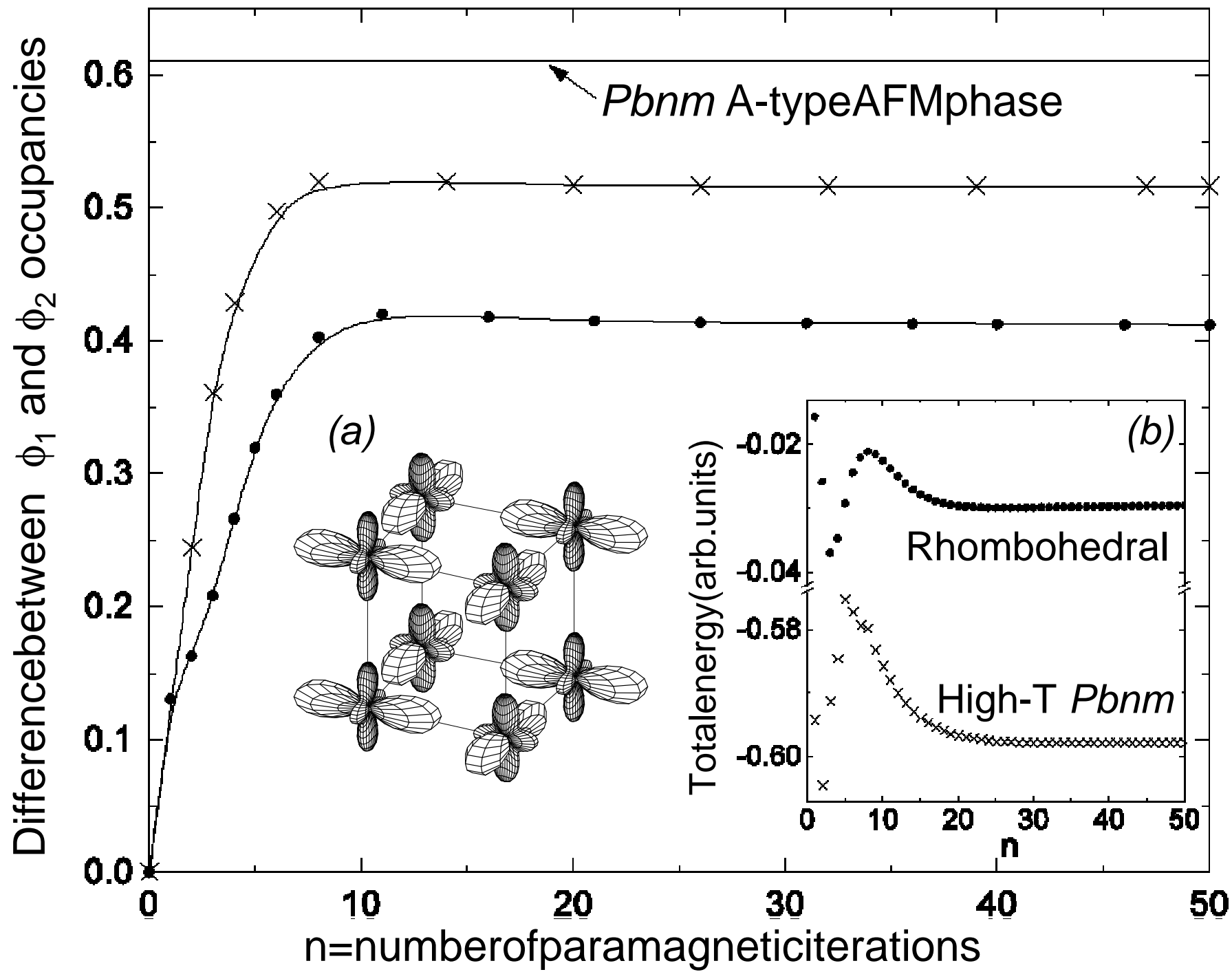
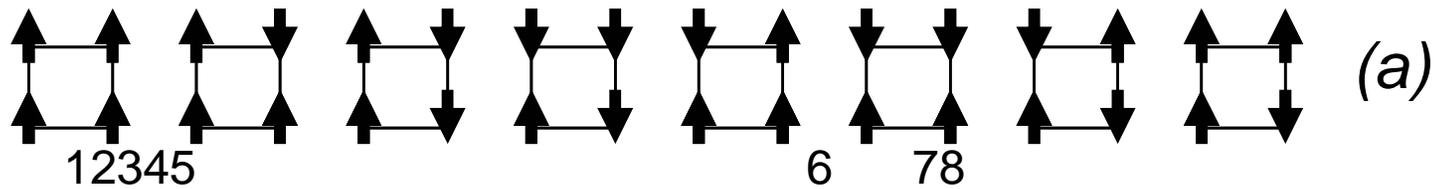
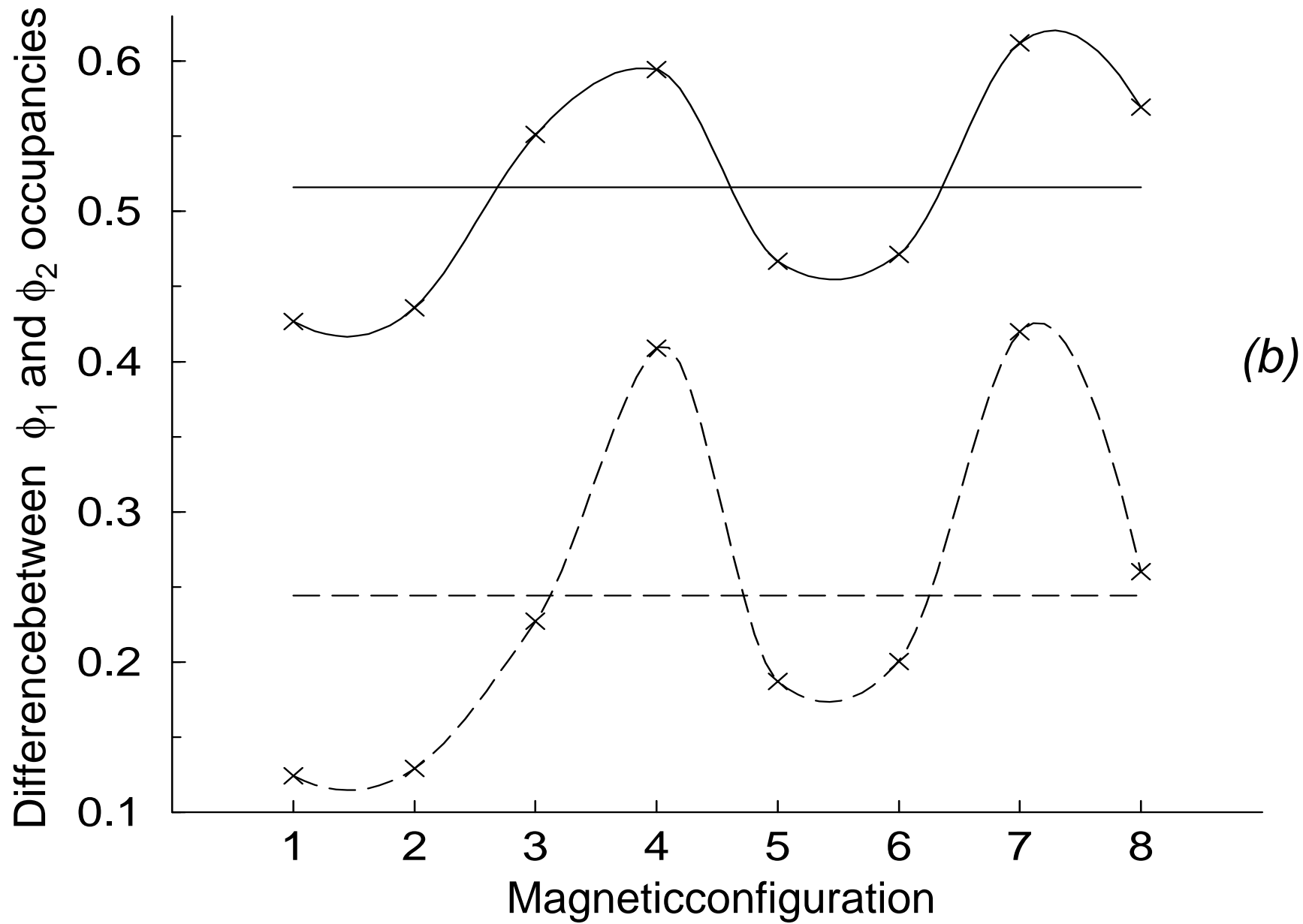
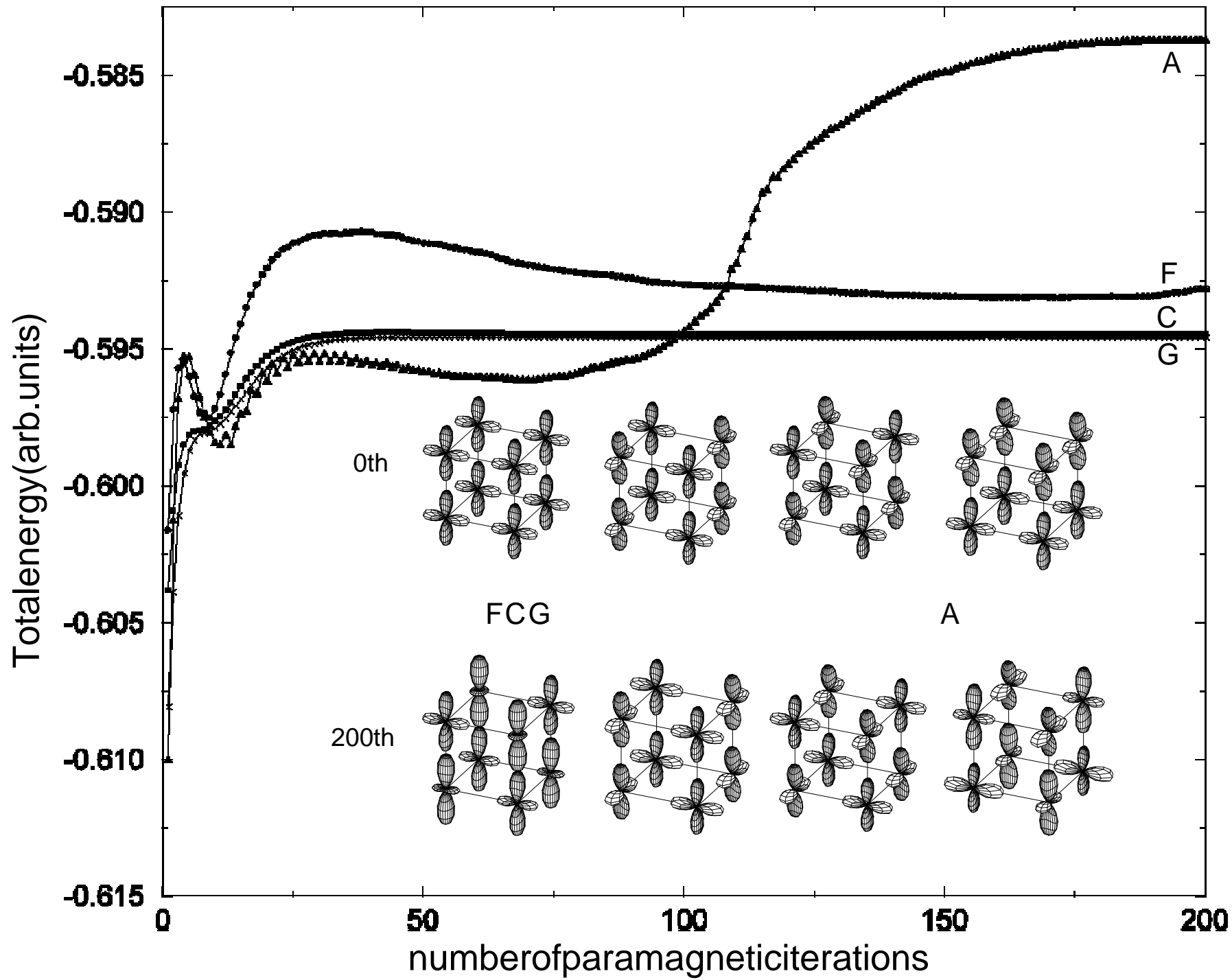
\end{document}